\documentclass[fleqn,usenatbib]{mnras}

\usepackage{newtxtext,newtxmath}

\usepackage[T1]{fontenc}
\usepackage{ae,aecompl}

\usepackage{graphicx}	
\usepackage{amsmath}	
\usepackage{amssymb}	
\usepackage{fixltx2e}
\usepackage{wasysym}
\usepackage{bold-extra}
\usepackage{longtable}
\usepackage{pdflscape}
\usepackage{afterpage}
\usepackage{epstopdf}

\newcommand{\sion}[2]{#1$\;$\textsc{#2}\relax}
\newcommand{\sub}[2]{\ifmmode #1_\mathrm{\scriptstyle #2} \else $#1_\mathrm{\scriptstyle #2}$\fi}
\newcommand{\ssub}[2]{\ifmmode #1_\mathrm{\scriptscriptstyle #2} \else $#1_\mathrm{\scriptscriptstyle #2}$\fi}
\newcommand{\pme}[2]{$^{+#1}_{-#2}$}
\newcommand{\pe}[1]{$^{+#1}$}
\newcommand{\me}[1]{$_{-#1}$}

\title[Quasar Absorption Outflows]{The Contribution of Quasar Absorption Outflows to AGN Feedback}

\author[Miller et al.]{
Timothy R. Miller,$^{1}$
Nahum Arav,$^{1}$\thanks{E-mail: arav@vt.edu}
 Xinfeng Xu$^{1}$
and Gerard A. Kriss$^{2}$
\\
$^{1}$Department of Physics, Virginia Tech, Blacksburg, VA 24061, USA\\
$^{2}$Space Telescope Science Institute, 3700 San Martin Drive, Baltimore, MD 21218, USA\\
}

\date{Accepted XXX. Received YYY; in original form ZZZ}

\pubyear{2020}

\begin{document}
\label{firstpage}
\pagerange{\pageref{firstpage}--\pageref{lastpage}}
\maketitle

\begin{abstract}
Determining the distance of quasar absorption outflows from the central source ($R$) and their kinetic luminosity (\sub{\dot{E}}{k}) are crucial for understanding their contribution to active galactic nucleus (AGN) feedback. Here we summarize the results for a sample of nine luminous quasars that were observed with the Hubble Space Telescope. We find that the outflows in more than half of the objects are powerful enough to be the main agents for AGN feedback, and that most outflows are found at $R$ $>$ 100 pc. The sample is representative of the quasar absorption outflow population as a whole and is unbiased towards specific ranges of $R$ and \sub{\dot{E}}{k}. Therefore, the analysis results can be extended to the majority of such objects, including broad absorption line quasars. We find that these results are consistent with those of another sample (seven quasars) that is also unbiased towards specific ranges of $R$ and \sub{\dot{E}}{k}. Assuming that all quasars have absorption outflows, we conclude that most luminous quasars produce outflows that can contribute  significantly to AGN feedback. We also discuss the criterion for whether an outflow is energetic enough to cause AGN feedback effects.
\end{abstract}




\section{Introduction}\label{sec:int}
Quasar spectra show outflowing material along the line of sight that is propagating from the centers of quasars as blueshifted absorption troughs relative to the rest frame of the host quasar. Upward of 40\% \citep[][]{hew03,dai08,gan08,kni08} of the quasar population contains absorption outflows. These outflows are candidates for producing major feedback processes within active galactic nuclei (AGNs), which include restricting the host galaxy growth \citep[e.g.,][]{cio09,hop09,fau12,zub14,sch15,cho17,pei17,val20}, explaining the mass correlation between the central black hole and the galaxy's bulge \citep[e.g.][]{sil98,bla04,hop09,ost10,dub14,ros15,vol16,ang17,yua18,nom20}, and chemical enrichment of the intracluster and intergalactic medium \citep[ICM, IGM; e.g.,][]{sca04,kha08,tor10,bar11,tay15,tho15,bar18}.

Quasar outflows come in many flavors: molecular \citep[e.g.,][]{kri11,cic14,fer15}, atomic \citep[e.g.,][]{kan08,adi18,mor18}, ionized seen in emission \citep[e.g.,][]{rup13,har14,bis17}, ionized seen in absorption \citep[e.g.,][]{kor08,dun10,she11} and ionized seen in X-ray \citep[e.g.,][]{cha09,nar15,tom15} spectra. A good comparison between the different outflow manifestations is given in \citet{fio17}. See also Figure~\ref{fig:fiorefig} here adapted from that paper.

Our aim in this paper is to determine whether absorption outflows seen in the rest-frame UV of luminous quasars (with bolometric luminosity \sub{L}{bol} $>$ 10$^{46}$~erg~s$^{-1}$) are capable of producing AGN feedback effects. The potential quasar outflows have to produce the above mentioned feedback processes is directly determined by their kinetic luminosity (\sub{\dot{E}}{k}) \citep[e.g.,][]{sca04,hop10}. \sub{\dot{E}}{k} is proportional to both the distance from the central source ($R$) and the total outflowing column density (\sub{N}{H}). Therefore, we need to find a sample of outflows where both $R$ and \sub{N}{H} can be measured and where the sample is representative of the majority of objects showing absorption outflows.

Measuring \sub{N}{H}: Ionization equilibrium in quasar outflows is dominated by photoionization, where the outflow is characterized by its ionization parameter (\sub{U}{H}) and \sub{N}{H}. Usually, the spectral synthesis code Cloudy \citep[][]{fer17} is used to produce photoionization simulations \citep[e.g.,][]{ara13} for a range of \sub{U}{H} and \sub{N}{H} values. These simulations predict the column density of each ion (\sub{N}{ion}) in the outflow. The best \sub{U}{H} and \sub{N}{H} solution is the one where the predicted \sub{N}{ion} values best fit the measured \sub{N}{ion} from the absorption troughs of the outflow.

Measuring $R$: The most robust way to determine $R$ for quasar absorption outflows is by using troughs from ionic excited states. The column density ratio between the excited and resonance states yields the electron number density of the outflow. Combined with a knowledge of the outflow’s \sub{U}{H}, $R$ can be determined \citep[e.g.,][]{ara18}. Three subtypes of absorption outflows are used to determine $R$:

1. The majority of $R$ determinations arise from singly ionized species \citep[mainly from \sion{Fe}{ii} and \sion{Si}{ii}, e.g.,][]{dek01,ham01,moe09,aok11,luc14,cho20}. However, most outflows show absorption troughs only from more highly ionized species. Therefore, the applicability of $R$ derived from singly ionized species to the majority of outflows is model-dependent \citep[see discussion in Section 1 of ][]{dun10}. Furthermore, this is a heterogeneous sample where many of the objects were selected for unique features and/or indications that $R$ and/or \sub{\dot{E}}{k} would have large values. Therefore, this is not a representative sample even for the low ionization outflows.

2. For ground-based observations, the main high-ionization species with a measurable trough arising from an excited state is \sion{S}{iv}, which has resonance and excited level transitions at 1062.66 and 1072.97~\AA, respectively. Therefore, $R$ determinations for high-luminosity quasars using high-ionization diagnostics from the ground concentrated on using the above \sion{S}{iv} diagnostic \citep[e.g.,][]{bor12,bor13,cha15b,xu18}. Of special interest is the seven quasar sample of \citet{xu19}, which was designed to be unbiased
towards particular $R$ ranges (elaboration on this sample is given in Section~\ref{ssec:compsamp}).

3. Using the Hubble Space Telescope/Cosmic Origin Spectrograph (HST/COS), the 500--1050~\AA\ rest-frame region (hereafter EUV500) can be observed in quasars at redshift $z\sim1$. The EUV500 contains an order of magnitude more diagnostic troughs \citep[see Figure 1 in ][]{ara20} than are available in ground-based data, which cover the $\lambda>$ 1050~\AA\ rest-frame region. These include troughs from very high-ionization (VHI) species (ions with an ionization potential (IP)
above 100 eV: e.g., \sion{Ne}{viii}, \sion{Na}{ix}, \sion{Mg}{x}, and \sion{Si}{xii}) whose ionization phase carries most of the outflowing \sub{N}{H} \citep[e.g., ][]{ara13}. Recently, we observed a sample of 10 EUV500 objects that show outflows. As described in \citet{ara20} (and summarized in Section~\ref{sec:euv500} here), this sample is the most suitable to learn about the properties of absorption outflows in high luminosity quasars. First, these objects are representative of the majority of absorption outflows. Second, they are unbiased towards specific $R$, \sub{N}{H}, and velocity ranges. Third, they give a census of the dominant very high-ionization phase (VHP) described above. In this paper we summarize the results of this sample and compare them to the \sion{S}{iv} sample of \citet{xu19} as well as to other types of quasar outflows.

We note that the value of $R$ is also important to the feasibility of AGN feedback for another reason. $R$ is a direct measure of how widespread the impact of the outflow is. If the outflows are confined to the AGN nuclear region, they may not do much to impact gas accretion and star formation throughout the galaxy as a whole. As we show below, most of the outflows in our sample have $R$ of several hundred to a few thousand parsecs. That is, they are on a galactic scale.

This paper is structured as follows. Section~\ref{sec:euv500} describes how the EUV500 sample  was obtained, discusses which outflow properties are unbiased and describes the  relationship between this sample and the population of quasar absorption outflows as a whole.
Section~\ref{sec:res} starts by describing two comparison samples. It then details the physical properties, distances, and energetics of both the individual outflows and host quasars in the EUV500 sample as well as compares the samples. 
In the Discussion (Section~\ref{sec:dis}) we: a) advocate that the criterion for 
whether an outflow is energetic enough to cause AGN feedback effects should be based on the ratio $\frac{\dot{E}_{k}}{L_{\text{Edd}}}$ (where $L_{\text{Edd}}$ is the Eddington luminosity) and not of 
$\frac{\dot{E}_{k}}{L_{\text{bol}}}$; 
b) Extrapolate the results of the sample  to the majority of luminous quasars; and c) discuss the full census of $\dot{E}_{k}$. We summarize our results in section 5. 

\section{The EUV500 Sample: Characteristics and Results}\label{sec:euv500}
The EUV500 sample gets its name from the fact that the observations primarily cover portions of the EUV500 for each quasar. Distances are determined from excited state transitions from various ionic species, but mostly from \sion{Ne}{v} and \sion{Ne}{vi}.


The sample is comprised of nine quasars that were first observed in programs where the scientific goals were to use the quasar light to probe intervening absorption from the IGM, the circumgalactic medium (CGM), galaxy halos, or high-velocity clouds. Eight of the nine quasars are from the aforementioned dedicated survey for quasar outflows observed during program GO-14777 (PI: N. Arav). The ninth is HE 0238-1904, whose rest-frame spectra cover the EUV500 and was observed during program GO-11541 (PI: J. Green). We note that there were two additional quasars observed during program GO-14777 that are not included in this sample. The first, 2MASS J1436+0727, has one outflow and it shows only VHP troughs, which do not yield density/distance diagnostics. Therefore, its $R$ and $\sub{\dot{E}}{k}$ are undetermined, and we opted to remove this object from the sample. The second, LBQS 1206+1052, has a small redshift such that the rest-frame spectra covers primarily lambda > 1050 A and therefore is part of the main comparison sample discussed in section 3.1.

The details of the selection criteria for program GO-14777 are given in \citet{ara20} and summarized here. The redshift range was restricted to 0.5--1.5, and a minimum continuum flux of $2\times10^{-15}$ erg cm$^{-2}$ s$^{-1}$ \AA$^{-1}$ was required for each quasar. An outflow was identified by matching at least two troughs with the same velocity that arise from resonance transitions of either the high-ionization phase (HP; e.g., \sion{O}{iv}, \sion{N}{iv}, and \sion{S}{iv}) or the VHP (i.e., \sion{Ne}{viii} and \sion{Mg}{x}). This identification scheme prevented biases towards a particular phase (either phase was chosen), a particular $R$ scale (searched for only resonance lines), or a particular velocity (identified outflows at any velocity). 

These outflows do not show troughs from abundant singly ionized species with strong lines (e.g., \sion{N}{ii}, \sion{O}{ii} and \sion{S}{ii}), and therefore would not show troughs from \sion{Mg}{ii}, \sion{Si}{ii} or \sion{Fe}{ii} if observed 
at rest-frame wavelengths ($\lambda_{\rm rest}$) greater than 1050 \AA.
Thus, these are not low-ionization outflows.
In contrast, All nine objects have outflows that show throughs from high ionization species (e.g., \sion{O}{iv}, \sion{Ne}{v}). As detailed in \citet{ara20} section 4.1, 
such outflows would have a detectable \sion{C}{iv} $\lambda$1549 Å trough, 
labeling them as high-ionization outflows if observed only
at $\lambda_{\rm rest} > 1050$ \AA.  Therefore, our EUV500 sample is representative
of high ionization outflows, which are the large majority of observed quasar outflows.
We note that due to their relatively low redshift ($z<1.5$) the SDSS spectra of these objects do not cover the spectral region of the \sion{C}{iv} $\lambda$1549 Å trough.  However, the  SDSS wavelength range does cover low ionization species like  \sion{Mg}{ii}, \sion{Fe}{ii} and sometimes \sion{Al}{iii} for the EUV500 sample.  As expected, we do not see the absorption features for these low ionization species.

Detailed analysis of each outflow appears in \citet{ara13}, \citet{mil20a,mil20b,mil20c}, and \citet{xu20a,xu20b,xu20c}. \citet{ara20} give an example of how $R$ and 
$\sub{\dot{E}}{k}$ are extracted from the data and discuss several issues related to the EUV500 outflows: the many advantages of studying quasar outflows using
EUV500 data, including: a) measuring the dominant VHP of the outflow, b) determining the total $N_H$ and ionization structure of the
outflows, and c) outflow distance determinations; comparison with X-Ray observations of Seyfert and quasar outflows; comparison with earlier EUV500 observations of quasar
outflows; and Broad Absorption Line (BAL) definition for the EUV500.

Here we give a summary of the results for the whole sample. First, in the Appendix we give the derived results for both individual outflows (Table 1: Velocities, Velocity Widths, and Distances of Each Outflow System) and for the host quasar (Table 2: Quasar Properties and Total Outflow Energetics). These tables also contain the results for our main comparison sample (the \sion{S}{iv} sample). Second, our figures show some of the results graphically, while comparing them to those in the comparison samples.

\section{Comparison with other Samples }\label{sec:res}

\subsection{Two Comparison Samples}\label{ssec:compsamp}
The main comparison sample is a collection of seven quasars and is known in this work as the \sion{S}{iv} sample since all of the outflow distances are determined from the excited and resonance state transitions of \sion{S}{iv}, specifically \sion{S}{iv*} 1072.97 \AA\ and \sion{S}{iv} 1062.66 \AA, respectively. The rest-frame wavelength range for the majority of this sample is between 1000 \AA\ and 8000 \AA.
 In all instances, the ratio of the column densities between \sion{S}{iv*} and \sion{S}{iv} were not known a priori, preventing an $R$ scale bias just like the EUV500 sample. Detailed analysis of each outflow appears in \citet{bor12,bor13}, \citet{cha15b}, \citet{mil18}, and \citet{xu18,xu19}.

The secondary comparison sample is from \citet{fio17}, where the entire collection consists of over 80 AGNs and includes molecular and ionized emission outflows in addition to absorption outflows observed in the ultraviolet and X-ray. 

\begin{figure}
	\includegraphics[trim=7mm 0mm 0mm 0mm,clip=true,scale=0.29]{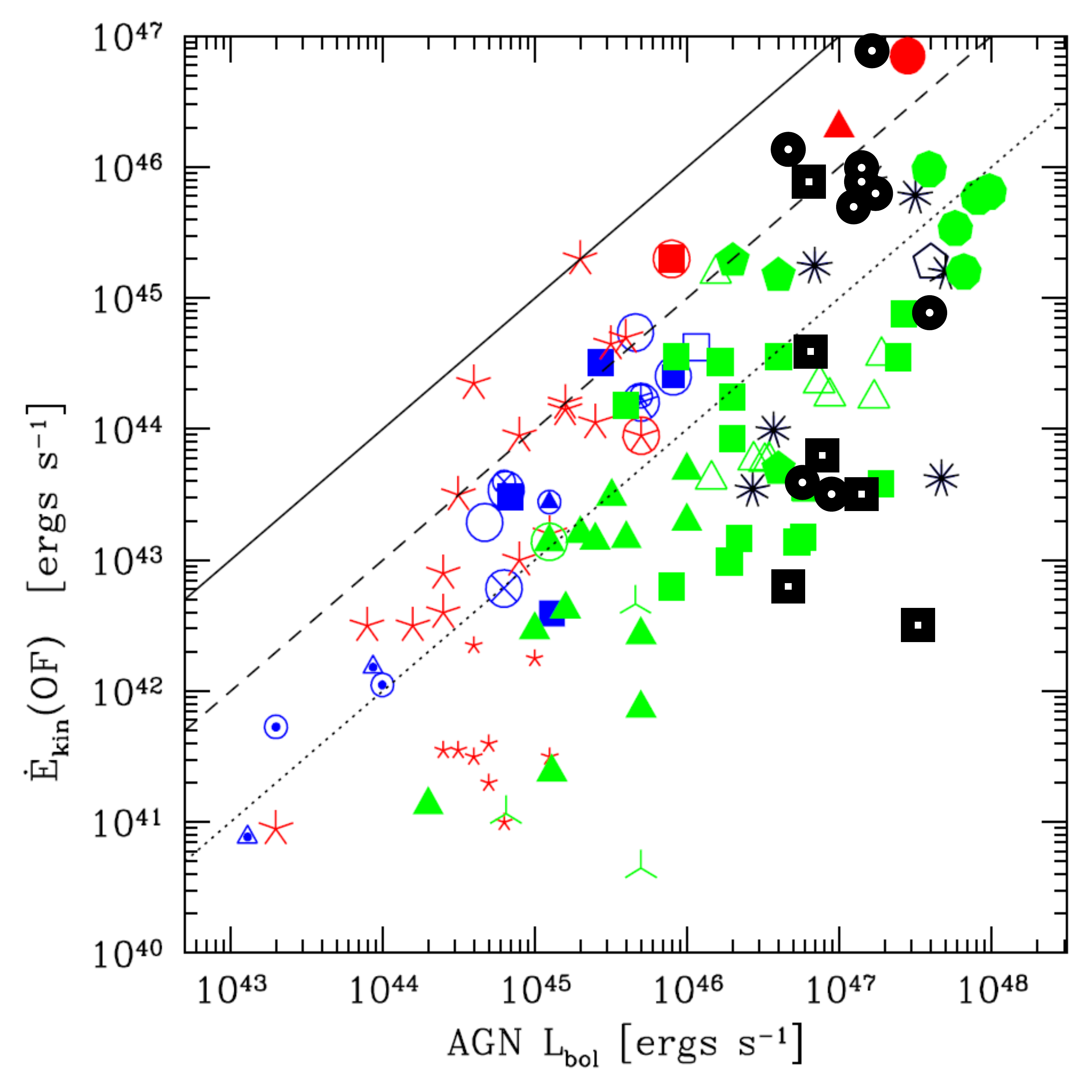}
	\caption{Right panel of Figure 1 from \citet{fio17} where \sub{\dot{E}}{kin}(OF) = \sub{\dot{E}}{k}. Molecular, ionized, and X-ray outflows are blue, green, and red respectively. The black stars are their BAL outflows. Our EUV500 and \sion{S}{iv} samples are overlaid in mostly filled, black circles and squares, respectively. The dotted, dashed, and solid lines denote where \sub{\dot{E}}{k} = 0.01, 0.1, 1.0 \sub{L}{bol}, respectively.}
	\label{fig:fiorefig}	
\end{figure}

\subsection{Comparison of All Samples}\label{ssec:comparisons}
For these three samples, the total \sub{\dot{E}}{k} for each quasar can be determined (see Section~\ref{sssec:qso} for the EUV500 and \sion{S}{iv} samples and Table B.1 of \citet{fio17} for the last sample) and compared. Figure~\ref{fig:fiorefig} makes such a comparison and shows the right panel of Figure 1 from \citet{fio17}, where overlaid on top in mostly filled black circles is the EUV500 sample and in mostly filled squares is the \sion{S}{iv} sample. 
As can be seen in figure 1 (see also the numerical values in table 2), the objects from both the EUV500 and the \sion{S}{iv} span the luminosity range $4\times10^{46}-4\times10^{47}$ erg s$^{-1}$. That is, they are all luminous quasars. This selection effect arises from the need to observe bright targets where a reasonable exposure time yield high enough signal-to-noise to enable the analysis. This is true for both the HST data of the EUV500 quasars at redshifts $0.5<z<1.5$, and for the \sion{S}{iv} quasars at redshifts $2<z<2.5$.  Six of the EUV500  and one of the \sion{S}{iv} objects comprise half of the 13 most energetic outflows shown in figure 1. The EUV500 object SDSS J1042+1646 has the largest \sub{\dot{E}}{k} across all samples.


\subsection{Comparison of the EUV500 and \sion{S}{iv} Samples}\label{ssec:twocomp}
Given the heterogeneous nature of the \citet{fio17} BAL outflows, the remaining comparisons and results are only on the EUV500 and \sion{S}{iv} samples.
\subsubsection{Individual Outflows}\label{sssec:io}
Table~\ref{tab:outflows} in the Appendix contains the velocities, velocity widths, and distances of the individual outflow systems. The velocity centroid marks the deepest part of the troughs associated with a given outflow system. The velocity of the widest trough is the midpoint of said trough (or blend of troughs between outflow systems) where continuous absorption below a residual intensity of 0.9 is observed. The width of the widest trough is for a single transition from the ion listed in the table and classifies each outflow as either a BAL or mini-BAL \citep[][]{ara20}. Only the systems with $R$ constraints are listed.

\begin{figure}
	\includegraphics[trim=13mm 25mm 28mm 45mm,clip=true,scale=0.30]{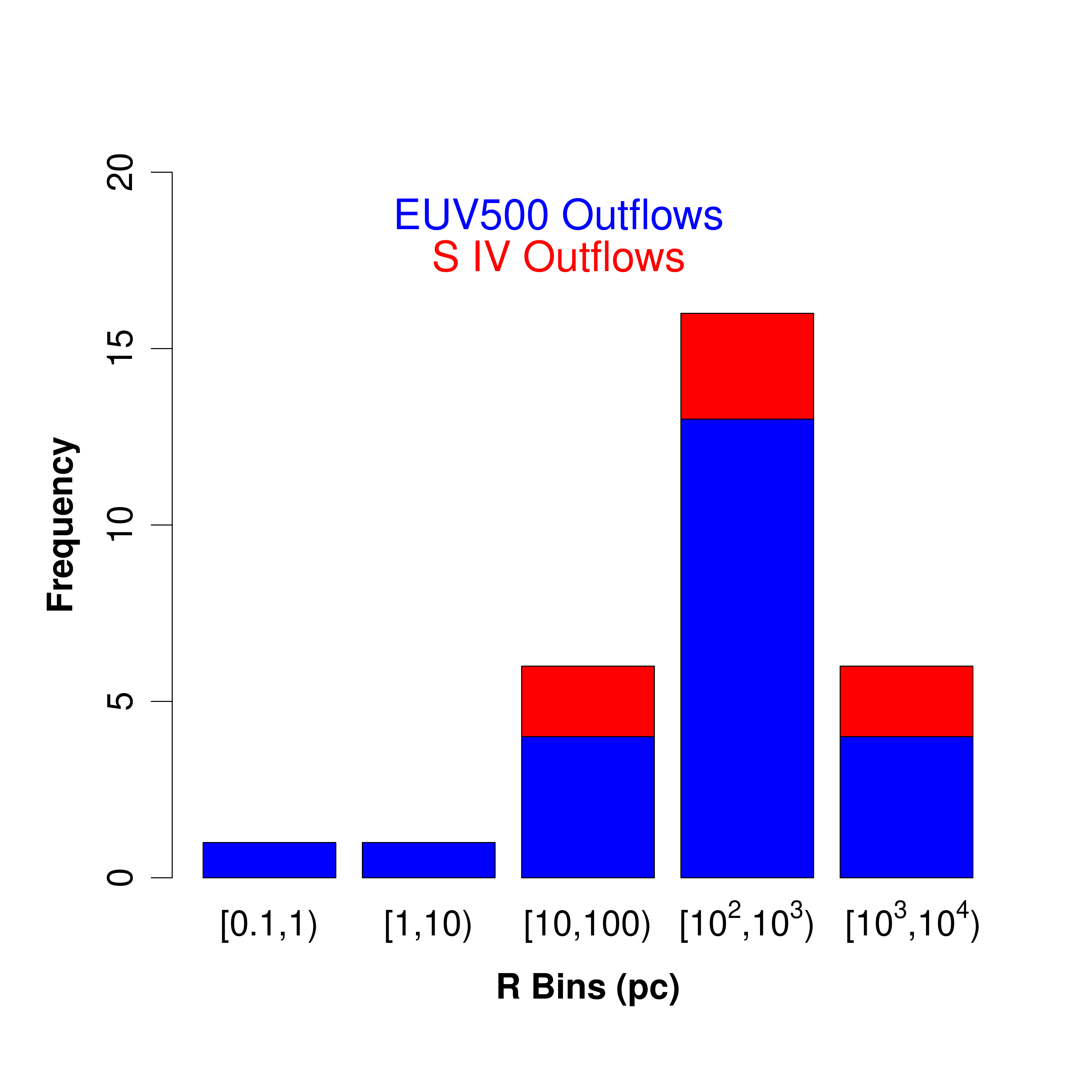}
	\caption{The distribution of distances ($R$) for the outflows where [x,y) = x $\le R <$ y. The majority of $R$ lie between 100 and 1000 pc.}
	\label{fig:R}	
\end{figure}

The distribution of $R$ is shown in Figure~\ref{fig:R} for the individual outflow distances listed in Table~\ref{tab:outflows}. We note that the $R$ distribution for the two independent samples are consistent with each other.
The distances for outflow systems 4 in UM 425 and 1 in SDSS J1135+1615 are excluded since they are upper limits. Also, only the lower limits for the distances of outflow systems 1 in VV2006 J1329+5405 and 2 in SDSS J1512+1119 are included. The majority of $R$ lie above 100 pc for both samples, and the maximum and minimum distances (excluding upper/lower limits) for the samples are 3400 pc and 0.15 pc, respectively.


\subsubsection{Host Quasars}\label{sssec:qso}
Table~\ref{tab:qsoprop} in the Appendix contains \sub{L}{bol}, central black hole mass (\sub{M}{BH}), and Eddington ratio ($\frac{L_{\text{bol}}}{L_{\text{Edd}}}$) of each quasar, obtained from the listed references. We note that the \sub{M}{BH} for HE 0238-1904 was calculated in this work by using the \sion{Mg}{ii}-based black hole mass equation from \citet{bah19}. Following their methodology, we measured the \sion{Mg}{ii} FWHM and local continuum level from Figure 1 of \citet{muz12}. The dominant uncertainty in all \sub{M}{BH} values is a systematic uncertainty of about 0.3 dex.

Based on the velocity width of the widest trough (see Table 1), for the EUV500 sample, five objects are classified as BALQSOs, while four are classified as mini-BALs.
For the \sion{S}{iv} sample five are BALQSOs and two are mini-BALs.

\begin{figure}
	\includegraphics[trim=11mm 20mm 25mm 39mm,clip=true,scale=0.40]{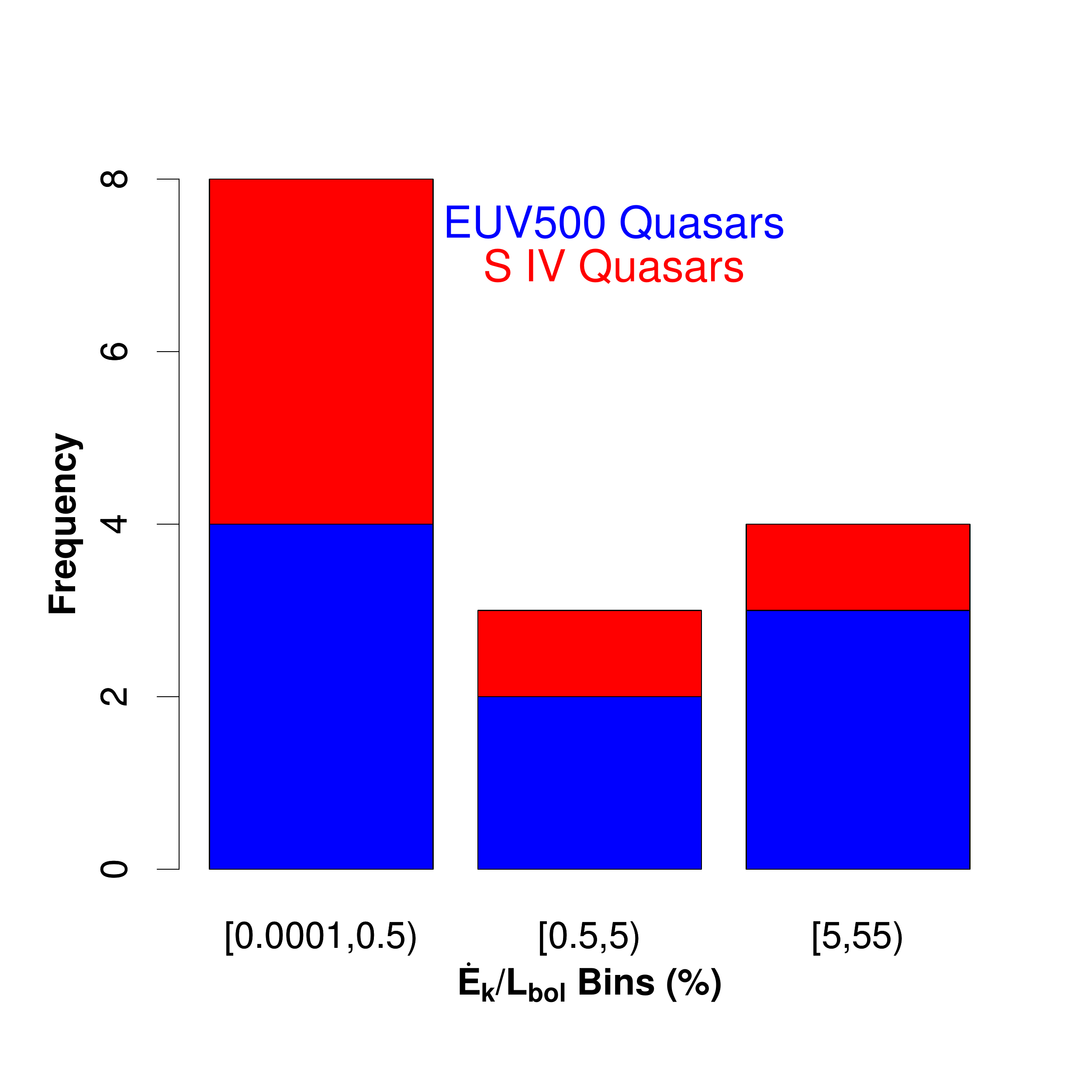}
	\caption{The distribution for the ratio of the total kinetic luminosity ($\dot{\sub{E}{k}}$) with respect to the bolometric luminosity (\sub{L}{bol}) for the quasars where [x,y) = x $\le \frac{\dot{E}_{k}}{\sub{L}{bol}}$ (\%) $<$ y.}
	\label{fig:edotlbol}	
\end{figure}

\begin{figure}
	\includegraphics[trim=12mm 25mm 33mm 47mm,clip=true,scale=0.3]{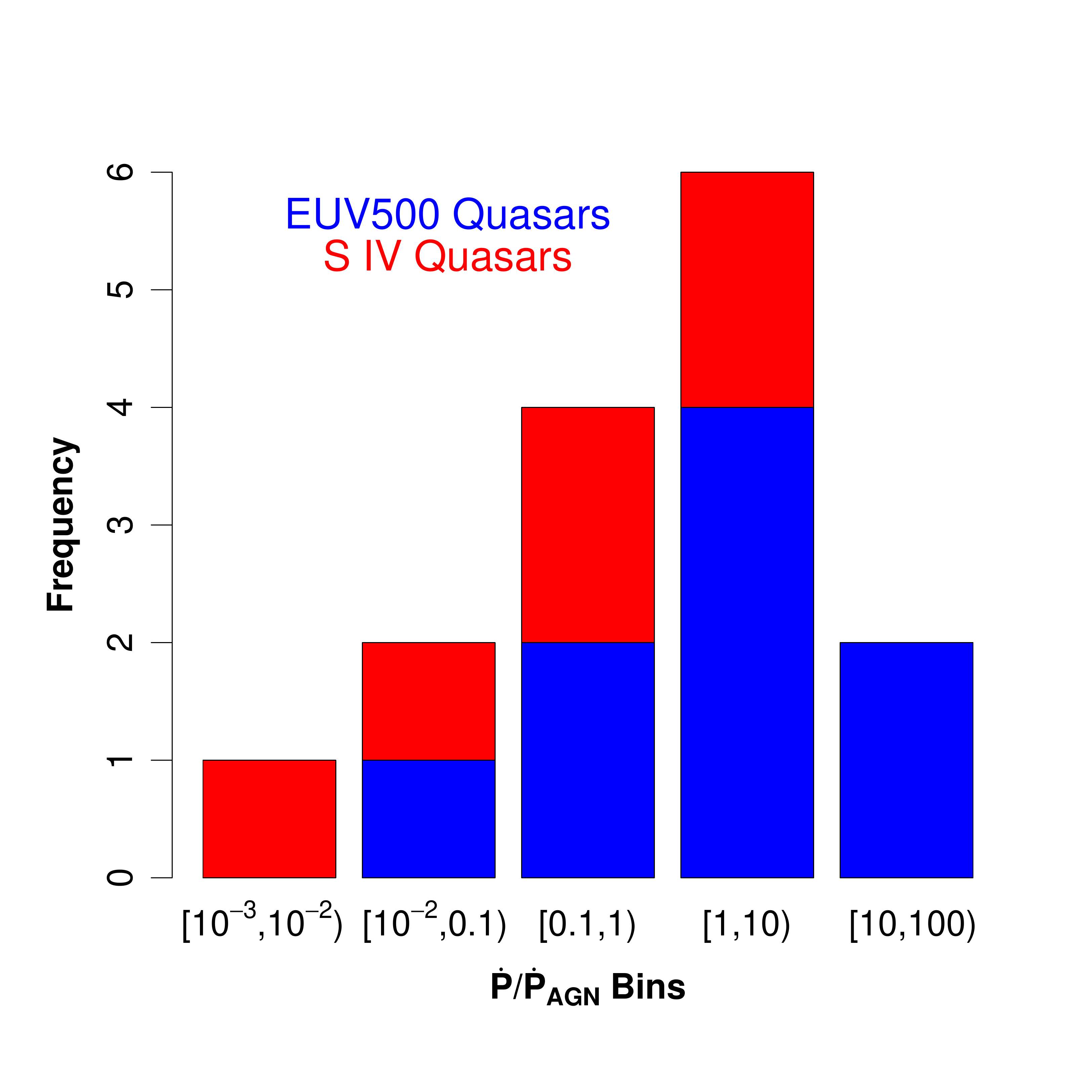}
	\caption{The distribution for the ratio of the total momentum flux ($\dot{P}$) with respect to the radiation momentum flux of the black hole ($\dot{P}_{AGN}$) for the quasars where [x,y) = x $\le \frac{\dot{P}}{\dot{P}_{AGN}} <$ y.}
	\label{fig:pdotpagn}	
\end{figure}

The total outflow energetics, i.e., the sum of the individual outflow energetics (excluding outflows with only upper limits) are also listed in Table~\ref{tab:qsoprop}. This includes the mass flux ($\dot{M}$), momentum flux ($\dot{P}$), and kinetic luminosity ($\sub{\dot{E}}{k}$). Note that, for consistency, we use the \sub{M}{BH} determined by \citet{xu19} for SDSS J0831+0354 instead of \citet{cha15b} in the calculation of $\frac{\dot{E}_{k}}{L_{\text{Edd}}}$.

The distribution for the ratio of $\sub{\dot{E}}{k}$ with respect to \sub{L}{bol} is shown in Figure~\ref{fig:edotlbol}. This ratio is commonly used within the literature when investigating the feedback potential of outflows \citep[e.g., ][]{cre12,har18}. However, in Section~\ref{ssec:lbvle} we argue that the ratio $\frac{\dot{E}_{k}}{L_{\text{Edd}}}$ is the meaningful physical comparison for assessing AGN feedback potential. The divisions at 0.5\% and 5\% arise from theoretical predictions by \citet{hop10} and \citet{sca04}, respectively, where those values mark the minimum ratio required for significant AGN feedback. Nearly half of the quasars have a ratio meeting at least one of these two thresholds. We note that this distribution does not change when \sub{L}{Edd} is substituted in for \sub{L}{bol}. This occurs since most of the quasars have \sub{L}{bol} within a factor of two of \sub{L}{Edd}, and the bins span an order of magnitude or more.  

Figure~\ref{fig:pdotpagn} shows the distribution of $\frac{\dot{P}}{\dot{{P}}\ssub{}{AGN}}$ (total momentum load) where $\dot{P}\ssub{}{AGN} = \sub{L}{bol}/c$ is the radiation momentum flux of the quasar. Momentum conserving outflows would have a value of 1 \citep[][]{fio17}. As can be seen, the majority of the EUV500 sample have values above 1 whereas the majority of the \sion{S}{iv} sample have values below 1.
The lower values (on average) of $\dot{P}$ and $\dot{E}_{k}$ for the \sion{S}{iv} sample compared to the EUV500 one are plausibly because the dominant outflow component measured by the VHP lines isn't measured in the \sion{S}{iv} sample (see section 4.3 for elaboration).


\section{Discussion}\label{sec:dis}

\subsection{The Energy Criterion for Outflows Influencing Feedback}\label{ssec:lbvle}

As noted in the previous section, several works use a specific percentage of the ratio $\frac{\dot{E}_{k}}{L_{\text{bol}}}$ as a threshold for whether an outflow is energetic enough to cause AGN feedback effects.
Here we make the case that the specific percentage should be of the ratio
$\frac{\dot{E}_{k}}{L_{\text{Edd}}}$ and not of 
$\frac{\dot{E}_{k}}{L_{\text{bol}}}$. For luminous quasars where $L_{\text{bol}}\sim L_{\text{Edd}}$ (as is the case for our samples) this point wouldn't matter much.  However, for Seyfert galaxies where the Eddington ratio $\frac{L_{\text{bol}}}{L_{\text{Edd}}}<0.1$, the difference will be more than an order of magnitude.
For example, using the criterion $\frac{\dot{E}_{k}}{L_{\text{bol}}}>0.5\%$,  \citet{cre12} find that 30\% of their Syfert outflows have enough energy to produce AGN feedback.  However, if instead $\frac{\dot{E}_{k}}{L_{\text{Edd}}}>0.5\%$ is used, none of these Seyferts have enough energy to produce AGN feedback.

 We outline our rationale for why a comparison with $L_{\text{Edd}}$ is more fundamental by first showing that  theoretical works that attempt to derive this ratio are often assuming that  
$L_{\text{bol}}=L_{\text{Edd}}$, and therefore their 
$\frac{\dot{E}_{k}}{L_{\text{bol}}}$ fraction is implicitly an 
$\frac{\dot{E}_{k}}{L_{\text{Edd}}}$ fraction.
One example is \citet{sca04} who state:
``The greatest uncertainty is 
$\epsilon_k\equiv\frac{\dot{E}_{k}}{L_{\text{bol}}}$, the fraction of the total bolometric luminosity (assumed to be the Eddington
luminosity) that appears as kinetic luminosity." Therefore, their $\epsilon_k$ is actually $\frac{\dot{E}_{k}}{L_{\text{Edd}}}$.  They find that when 
$\epsilon_k\geq5\%$ strong feedback effects are possible.

Another influential theoretical paper is \citet{hop10} which gives 
 the lower limit for the amount of energy needed for AGN feedback.  
In the abstract they write:
"$\sim$0.5 per cent of the luminosity" and  the question is whether "luminosity" here refers to the $L_{\text{bol}}$ or  $L_{\text{Edd}}$.  The relevant portion of the paper is:

``Another way of stating this is,
for accretion with an Eddington ratio $\dot{m}$  and black hole mass 
$M_{\rm BH}$ relative
to the expectation $\langle M_{\rm BH}\rangle$ from the $M_{\rm BH}$–$\sigma$ relation, the relevant
outflows will be driven (and star formation suppressed) when
$$\eta \dot{m}\frac{M_{\rm BH}}{\langle M_{\rm BH}\rangle}\sim0.05f_{\rm hot} \ \ \ \ \ \ \ \ \ \ \ \ (9)$$
where $\eta$ is the feedback efficiency $(\dot{E}=\eta L)$''
(Note that $f_{\rm hot}\sim0.1$ is the mass fraction
in the hot diffuse ISM.)

By substituting $\eta=\dot{E}/L$ and $\dot{m}=L/L_{\rm Edd}$ into their equation (9), we obtain the criterion $$\frac{\dot{E}}{L_{\rm Edd}}\frac{M_{\rm BH}}{\langle M_{\rm BH}\rangle}\sim0.05f_{\rm hot}$$ which clearly depends on the fraction 
$\frac{\dot{E}_{k}}{L_{\text{Edd}}}$ and not on 
$\frac{\dot{E}_{k}}{L_{\text{bol}}}$.

\citet{diM05Nat} state: ``We further assume that a
small fraction, $f$, of the radiated luminosity couples thermodynamically
to the surrounding gas."' However they also state:
“Owing to the enhanced gas density, the black holes, which also merge to form one object, experience a rapid phase of accretion close to the Eddington rate, resulting in significant mass growth.”  This suggests that they need $L_{\text{bol}}$ to be close to the Eddington rate. See also their figure 2.
\citet{sil98} describe in their Section 2.1 the mechanical (i.e. outflow) luminosity as a fraction of $L_{\text{Edd}}$ and detail the conditions required to expel gas from the host protogalaxy. 

Two physical considerations:\\
a) $L_{\text{bol}}$ (the total amount of electromagnetic energy emitted per unit of time) can vary by orders of magnitude over time in the same object. Therefore, if it is $\frac{\dot{E}_{k}}{L_{\text{bol}}}$ that is important, the ratio will vary similarly with changes of $L_{\text{bol}}$.  This does not seem to be physically plausible as the feedback effects are connected to the (slowly changing) parameters of the host galaxy. \\
b) Physically, AGN feedback works if a fraction of the rest mass energy of the forming black hole is entrained back in the host galaxy \citep[e.g.,][]{Loeb05}.  This fraction is related to $L_{\text{Edd}}$ and not to $L_{\text{bol}}$, which can be arbitrarily smaller.

 Therefore, the potential for AGN feedback from an outflow should be conditional on a percentage of \sub{L}{Edd} rather than a percentage of \sub{L}{bol}.

\subsection{Extrapolation of the results to the majority of luminous quasars}\label{ssec:extrapolation2}  
In section 2.1 we demonstrated that  the EUV500 sample
is representative of high ionization outflows, which are the large
majority of observed quasar outflows. Here we argue that these results can be extended to the majority of all luminous quasars.

Quasar absorption outflows are seen in only a portion of all quasar spectra.  For example, high-ionization BALs are detected in roughly 20\% of all quasars \citep[e.g.,][]{hew03}.  The most common explanation for this percentage is that all quasars have BAL outflows, which cover only 20\% of the solid angle around the object \citep[e.g.,][]{wey91,rei03}.  Therefore, in only 20\% of the cases our line of sight towards the quasar intercepts a BAL outflow. Our calculation of the kinetic luminosity takes this consideration into account by multiplying the applicable result for the full solid angle by the appropriate detection-fraction for the studied population. That is, for high-ionization BALQSOs we multiply the full solid angle result by 20\% in order to derive the actual energy such an outflow has on average. This procedure assumes implicitly that such outflows reside in all luminous quasars.  
Assuming that all quasars have absorption outflows, we conclude that most luminous quasars produce outflows that can create significant AGN feedback.


\subsection{The Full Census of $\dot{E}_{k}$}\label{ssec:census}

Are we accounting for most of the $\dot{E}_{k}$ associated with all the outflows in a given object?  

For the EUV500 objects there are two issues that can cause underestimation of $\dot{E}_{k}$.  First, we have shown that the majority of the measured outflowing $N_H$ resides in the VHP.  However, our highest ionization potential (IP) diagnostic is \sion{Si}{xii} with IP=523 eV.  X-ray outflows in Seyfert galaxies (the so called warm-absorbers) show diagnostics at much higher energy (e.g., IP=9278 eV for \sion{Fe}{xxvi}).  Modeling such spectra show that the ionization phase above what can be probed with \sion{Si}{xii} often carries a larger amount of $N_H$ than the \sion{Si}{xii} and lower ionization phases \citep[e.g.,][]{kaa14}.  It is reasonable to assume that the absorption outflows in luminous quasars have the same ionization distribution as the warm absorbers. If that is the case, we are missing most of the $N_H$ of the EUV500 outflows.  Second, out of the 29 outflows we detect in the EUV500 survey, six are only detected by their VHP phase and they lack $R$ diagnostics \citep{ara20,mil20c}.  Therefore, their $\dot{E}_{k}$ are undetermined and are missing from the total $\dot{E}_{k}$ census.

The underestimation of $\dot{E}_{k}$ is probably larger for the \sion{S}{iv} objects. First, in these objects, we do not cover the spectral region of the VHP diagnostics (e.g., \sion{Ne}{viii} and \sion{Mg}{x}). Therefore, we cannot measure the VHP of the outflow which is shown to carry the majority of the outflowing $N_H$ 
in the EUV500 objects. Second, most \sion{S}{iv} objects show several outflows in  \sion{C}{iv}, but only one system where an \sion{S}{iv} trough is associated with the \sion{C}{iv} trough.  It is only that system from which we extract $\dot{E}_{k}$.   Each of the other \sion{C}{iv} systems has an $\dot{E}_{k}$ that we cannot determine.  Therefore our $\dot{E}_{k}$ estimate for the object is a lower limit to the total $\dot{E}_{k}$ in that object. 
Third, like in the case of the EUV500 objects, we may be missing material with even higher ionization than we can detect through the VHP.


\subsection{Correlations Between Outflow Properties}\label{ssec:compow}
As can be seen in Figure~\ref{fig:fiorefig}, the small dynamical range of our samples in \sub{L}{bol} prevents meaningful correlations to be made with the outflow properties (see also the left panels of Figures 1 and 2 in \citet{fio17} for other correlations they observed). However, we draw the same conclusions as \citet{cre12} about the correlations between individual outflow properties, namely there are no major  correlations (see their first four figures). 

\section{Summary}\label{sec:sum}
In this paper we summarize the results from our sample of EUV500 quasar absorption outflows and compared them with other samples.  Our main results are:

\begin{enumerate}

\item The outflows in more than half of the EUV500 objects are powerful enough to be the main agents for AGN feedback, and most of the outflows in these objects are found at $R$ $>$ 100 pc (see tables 1 and 2). 

\item The sample is representative of the quasar absorption outflow population as a whole. Specifically, it represents the high ionization outflows,
 which are the large majority of quasar UV absorption outflows. 
 Furthermore, the sample is unbiased towards specific ranges of $R$ and \sub{\dot{E}}{k}. Therefore, the analysis results can be extended to the majority of such objects, including broad absorption line quasars (see section \ref{sec:euv500}). 

\item We find that these results are consistent with those of another sample: seven ground based observed quasars, which show absorption troughs from \sion{S}{iv} and \sion{S}{iv}* transitions.  This comaparison sample is also unbiased towards specific ranges of $R$ and \sub{\dot{E}}{k} (see section \ref{ssec:twocomp}) .

\item We compare our results with a large heterogeneous sample of different types of AGN outflows (molecular, ionized seen in emission, ionized seen in absorption, and X-ray). Six of the EUV500
and one of the \sion{S}{iv} objects comprise half of the 13 most energetic
outflows shown in figure 1 
(see section \ref{ssec:comparisons}).

\item It is generally assumed that all quasars have absorption outflows, where the detection fraction of absorption outflows is similar to the fraction of solid angle subtended by the outflow around the source.  Under this assumption, 
we conclude that most luminous quasars produce outflows that can contribute  significantly to AGN feedback (see section \ref{ssec:extrapolation2}). 

\item We also discuss the criterion for whether an outflow is energetic enough to cause AGN feedback effects, and conclude that 
instead of using a specific percentage of the ratio $\frac{\dot{E}_{k}}{L_{\text{bol}}}$ the criterion should be based on a percentage of the ratio
$\frac{\dot{E}_{k}}{L_{\text{Edd}}}$   (see section \ref{ssec:lbvle}).

\end{enumerate}

\section*{Acknowledgements}

T.M., N.A., and X.X. acknowledge support from NASA  grants \textit{HST} GO 14777, 14242, 14054, and 14176 as well as \textit{HST} AR 15786. This support is provided by NASA through a grant from the Space Telescope Science Institute, which is operated by the Association of Universities for Research in Astronomy, Incorporated, under NASA contract NAS5-26555. T.M. and N.A. also acknowledge support from NASA ADAP 48020 and NSF grant AST 1413319.  We thank Tiago Costa and Chris Harrison for illuminating discussions. We thank Fabrizio Fiore for letting us use the right panel of figure 1 from his \citet{fio17} paper, as the base for our figure~1.

\section*{Data Availability Statement}
The data underlying this article are available in the article, especially in tables 1 and 2.








\section*{Appendix}
\begin{table*}
	\caption{Velocities, Velocity Widths, and Distances of Each Outflow System\label{tab:outflows}}
	\small
	\begin{tabular}{c c c c c c}
		\hline\hline
		System   & $v^{(a)}$ & $\sub{v}{wt}^{(b)}$ & $\Delta\sub{v}{wt}^{(c)}$ & $\mathrm{\sub{Ion}{wt}^{(d)}}$ & R \\
		 & (km s$^{-1}$) & (km s$^{-1}$) & (km s$^{-1}$) & & (pc)\\
		\hline
		\multicolumn{6}{c}{\textbf{EUV500 Outflows}} \\
		\hline
		\multicolumn{6}{l}{\textbf{HE 0238-1904}} \\
		\hline
		1 & -3850 & -3850 & 500 & \sion{Ne}{viii} & 2000\pme{1200}{320} \\
		2 & -5000 & -5000 & 500 & \sion{Ne}{viii} & 3400\pme{2000}{490} \\
		\hline
		\multicolumn{6}{l}{\textbf{PKS J0352-0711}} \\
		\hline
		1 & -1950 & -1950 & 500 & \sion{Ne}{viii} & 520\pme{300}{150} \\
		2 & -3150 & -3150 & 1400 & \sion{Ne}{viii} & 8.9\pme{4.9}{4.5} \\
		\hline
		\multicolumn{6}{l}{\textbf{SDSS J0755+2306}} \\
		\hline
		1 & -5520 & -5520 & 1800 & \sion{Ne}{viii} & 270\pme{100}{90} \\
		2 & -9660 & -9660 & 3200 & \sion{Ne}{viii} & 1600\pme{2000}{1100} \\
		\hline
		\multicolumn{6}{l}{\textbf{SDSS J0936+2005}} \\
		\hline
		1 & -7960 & -8100 & 750 & \sion{Ne}{viii} & 77\pme{40}{22} \\
		2 & -8200 & -8100 & 750 & \sion{Ne}{viii} & 150\pme{50}{50} \\
		3 & -9300 & -9300 & 500 & \sion{Ne}{viii} & 14\pme{9}{4} \\
		\hline
		\multicolumn{6}{l}{\textbf{SDSS J1042+1646}} \\
		\hline
		1 & -4950 & -4950 & 2500 & \sion{Ne}{viii} & 840\pme{500}{300} \\
		2 & -5750 & -5750 & 1500 & \sion{Ne}{viii} & 800\pme{300}{200} \\
		3 & -7500 & -7500 & 1350 & \sion{Ne}{viii} & 15\pme{8}{8} \\
		\hline
		\multicolumn{6}{l}{\textbf{2MASS J1051+1247}} \\
		\hline
		1 & -4900 & -5300 & 1200 & \sion{Ne}{viii} & 460\pme{200}{130} \\
		2 & -5150 & -5300 & 1200 & \sion{Ne}{viii} & 360\pme{130}{100} \\
		3 & -5350 & -5300 & 1200 & \sion{Ne}{viii} & 180\pme{220}{50} \\
		4 & -5650 & -5300 & 1200 & \sion{Ne}{viii} & 460\pme{160}{140} \\
		\hline
		\multicolumn{6}{l}{\textbf{UM 425}} \\
		\hline
		1 & -1640 & -1900 & 1600 & \sion{Mg}{x} & 1180\pme{430}{290} \\
		2 & -1980 & -1900 & 1600 & \sion{Mg}{x} & 760\pme{440}{320} \\
		3 & -2200 & -1900 & 1600 & \sion{Mg}{x} & 340\pme{370}{190} \\
		4 & -9420 & -9420 & 800 & \sion{Mg}{x} & $^{e}$$<$22\pe{37} \\
		\hline
		\multicolumn{6}{l}{\textbf{VV2006 J1329+5405}} \\
		\hline
		1 & -11600 & -12300 & 2800 & \sion{Ne}{viii} & $^{f}$0.15\me{0.11}---0.4\pe{17.4} \\
		\hline
		\multicolumn{6}{l}{\textbf{7C 1631+3930}} \\
		\hline
		1 & -1010 & -1400 & 1000 & \sion{Ne}{viii} & $^{g}$$>$19\me{4} \\
		2 & -1430 & -1400 & 1000 & \sion{Ne}{viii}	& 940\pme{260}{230} \\
		4 & -5770 & -5700 & 1100 & \sion{Ne}{viii}	& 590\pme{470}{270} \\
		\hline
		\multicolumn{6}{c}{\textit{Table~\ref{tab:outflows} continued}}\\
	\end{tabular}		
\end{table*}

\begin{table*}
	\small
	\begin{tabular}{c c c c c c}
		\multicolumn{6}{c}{\textbf{Table~\ref{tab:outflows}} (\textit{continued})}\\
	\hline\hline
	System   & $v^{(a)}$ & $\sub{v}{wt}^{(b)}$ & $\Delta\sub{v}{wt}^{(c)}$ & $\mathrm{\sub{Ion}{wt}^{(d)}}$ & R \\
	 & (km s$^{-1}$) & (km s$^{-1}$) & (km s$^{-1}$) & & (pc)\\
	\hline
	\multicolumn{6}{c}{\textbf{\sion{S}{iv} Outflows}}\\
	\hline
	\multicolumn{6}{l}{\textbf{SDSS J0046+0104}} \\
	\hline
	1 & -1730 & -1730 & 3700 & \sion{C}{iv} & 1200\pme{250}{450}\\
	\hline
	\multicolumn{6}{l}{\textbf{SDSS J0831+0354}} \\
	\hline
	1 & -10800 & -10800 & 5000 & \sion{C}{iv} & 78\pme{27}{18}\\
	\hline
	\multicolumn{6}{l}{\textbf{SDSS J0941+1331}} \\
	\hline
	1 & -3180 & -3180 & 8000 & \sion{C}{iv} & 200\pme{40}{60}\\
	\hline
	\multicolumn{6}{l}{\textbf{SDSS J1111+1437}} \\
	\hline
	1 & -1860 & -1860 & 1800 & \sion{C}{iv} & 880\pme{210}{260}\\
	\hline
	\multicolumn{6}{l}{\textbf{SDSS J1135+1615}} \\
	\hline
	1 & -7250 & -7250 & 9500 & \sion{C}{iv} & $^{e}$$<$40\pe{10}\\
	\hline
	\multicolumn{6}{l}{\textbf{LBQS J1206+1052}} \\
	\hline
	1 & -1400 & -1400 & 1050 & \sion{N}{v} & 500\pme{100}{110}\\
	\hline
	\multicolumn{6}{l}{\textbf{SDSS J1512+1119}} \\
	\hline
	1 & -1050 & -1450 & 2000 & \sion{C}{iv} & $^{g}$$>$3000\me{150}\\
	2 & -1850 & -1450 & 2000 & \sion{C}{iv} & $^{f}$10\me{0.5}---300\pe{15}\\
	\hline
	\multicolumn{6}{l}{Note:}\\
	\multicolumn{6}{l}{See Table~\ref{tab:qsoprop} for references.}\\
	\multicolumn{6}{l}{(a). The velocity centroid of each outflow system.}\\
	\multicolumn{6}{l}{(b). The velocity at the middle of the widest trough. Blended troughs}\\
\multicolumn{6}{l}{between outflows have the same value.}\\
\multicolumn{6}{l}{(c). Velocity width of the widest trough. Determined by continuous}\\
\multicolumn{6}{l}{absorption below a residual intensity of 0.9. Blended troughs between}\\
\multicolumn{6}{l}{outflows have the same value.}\\\multicolumn{6}{l}{(d). Ion of the widest trough.}\\
	\multicolumn{6}{l}{(e). Upper limit with a 1-$\sigma$ uncertainty.}\\
	\multicolumn{6}{l}{(f). Both a lower and upper limit with a 1-$\sigma$ uncertainty for each limit.}\\
	\multicolumn{6}{l}{(g). Lower limit with a 1-$\sigma$ uncertainty.}\\
	\end{tabular}		
\end{table*}

\begin{table*}
	\caption{Quasar Properties and Total Outflow Energetics\label{tab:qsoprop}}
	\small
	\begin{tabular}{l c c c c c c c c}
	\hline\hline
	Quasar & L$_{\text{bol}}$ & log($\mathrm{\sub{M}{BH}}$)$^{(a)}$ & $\frac{L_{\text{bol}}}{L_{\text{Edd}}}$ & $\dot{\text{M}}$ & $\mathrm{\frac{\dot{P}}{\dot{{P}}\ssub{}{AGN}}}$ & log($\sub{\dot{E}}{k}$) & $\frac{\dot{E}_{k}}{L_{\text{bol}}}$ & $\frac{\dot{E}_{k}}{L_{\text{Edd}}}$$^{(b)}$\\
	 & (10$^{47}$ erg s$^{-1}$) & log($\mathrm{\sub{M}{\astrosun}}$) & & (\sub{\text{M}}{\astrosun} yr$^{-1}$) & & log(ergs s$^{-1}$) & (\%) & (\%)\\
	\hline
	\multicolumn{9}{c}{\textbf{EUV500 Outflows}} \\
	\hline
	HE 0238-1904$^{(c)}$ & 1.6 & 10.0 & 0.13 & 190\pme{80}{150} & 1.1\pme{0.5}{0.9} & 45.8\pme{0.2}{0.9} & 3.9\pme{4.7}{3.1} & 0.51\pme{0.62}{0.41} \\
	PKS J0352-0711$^{(d)}$ & 0.55 & 8.9 & 0.54 & 19\pme{24}{10} & 0.16\pme{0.18}{0.08} & 43.6\pme{0.3}{0.3} & 0.067\pme{0.071}{0.033} & 0.034\pme{0.072}{0.013} \\
	SDSS J0755+2306$^{(e)(m)}$ & 0.44 & 8.9 & 0.43 & $>$450 & $>$19 & $>$46.1 & $>$30 & $>$13 \\
	SDSS J0936+2005$^{(f)}$ & 1.3 & 9.0 & 1.0 & 390\pme{280}{180} & 4.6\pme{3.4}{2.2} & 45.9\pme{0.2}{0.3} & 5.7\pme{8.5}{3.5} & 5.7\pme{8.5}{3.5} \\
	SDSS J1042+1646$^{(g)}$ & 1.5 & 9.3 & 0.59 & 7180\pme{1220}{1700} & 49\pme{10}{12} & 46.9\pme{0.1}{0.1} & 51\pme{21}{8.5} & 31\pme{23}{11} \\
	2MASS J1051+1247$^{(h)}$ & 1.3 & 9.0 & 1.0 & 1010\pme{640}{280} & 7.8\pme{5.0}{2.2} & 46.0\pme{0.2}{0.1} & 7.0\pme{6.6}{2.3} & 7.0\pme{6.6}{2.3} \\
	UM 425$^{(f)}$ & 3.8 & 9.7 & 0.59 & 1050\pme{680}{470} & 0.86\pme{0.57}{0.39} & 44.9\pme{0.3}{0.2} & 0.21\pme{0.21}{0.08} & 0.15\pme{0.22}{0.08} \\
	VV2006 J1329+5405$^{(f)(n)}$ & 0.89 & 9.0 & 0.70 & $>$0.60\me{0.40} & $>$0.015\me{0.010} & $>$43.4\me{0.6} & $>$0.028\me{0.020} & $>$0.022\me{0.017} \\
	7C 1631+3930$^{(f)}$ & 1.2 & 9.3 & 0.47 & 610\pme{520}{290} & 4.8\pme{4.8}{2.7} & 45.7\pme{0.3}{0.4} & 4.2\pme{4.2}{2.5} & 2.3\pme{4.9}{1.4} \\
	\hline
	\multicolumn{9}{c}{\textbf{\sion{S}{iv} Outflows}}\\
	\hline
	SDSS J0046+0104$^{(i)}$ & 1.3 & 9.0 & 1.0 & 37\pme{4}{3} & 0.096\pme{0.016}{0.013} & 43.5\pme{0.04}{0.04} & 0.031\pme{0.003}{0.003} & 0.031\pme{0.034}{0.014} \\
	SDSS J0831+0354$^{(j)}$ & 0.62 & 8.8 & 0.77 & 230\pme{330}{130} & 7.6\pme{11.1}{4.3} & 45.9\pme{0.4}{0.3} & 26\pme{39}{13} & 20\pme{50}{13} \\
	SDSS J0941+1331$^{(i)}$ & 0.63 & 8.8 & 0.78 & 120\pme{14}{13} & 1.1\pme{0.2}{0.2} & 44.6\pme{0.04}{0.05} & 0.63\pme{0.10}{0.09} & 0.50\pme{0.68}{0.29} \\
	SDSS J1111+1437$^{(k)}$ & 0.79 & 9.2 & 0.39 & 55\pme{10}{11} & 0.24\pme{0.05}{0.05} & 43.8\pme{0.1}{0.1} & 0.079\pme{0.017}{0.018} & 0.032\pme{0.044}{0.019} \\
	SDSS J1135+1615$^{(k)(o)}$ & 1.6 & 9.0 & 1.3 & $<$150\pe{10} & $<$1.3\pe{0.2} & $<$45.4\pe{0.03} & $<$1.6\pe{0.2} & $<$2.0\pe{2.7} \\
	LBQS J1206+1052$^{(l)}$ & 0.44 & 9.2 & 0.22 & 8.9\pme{7.2}{3.1} & 0.054\pme{0.045}{0.019} & 42.8\pme{0.3}{0.2} & 0.015\pme{0.012}{0.005} & 0.0028\pme{0.0042}{0.0016} \\
	SDSS J1512+1119$^{(i)(n)}$ & 3.2 & 9.1 & 2.0 & $>$4.4\me{0.2} & $>$0.0032\me{0.0003} & $>$42.4\me{0.01} & $>$0.0008\me{0.00007} & $>$0.0016\me{0.0006} \\
	\hline
	\multicolumn{6}{l}{Note:}\\
		\multicolumn{9}{l}{(a). Accurate to about 0.3 dex.}\\
		\multicolumn{9}{l}{(b). Includes the uncertainties in \sub{M}{BH}}\\
		\multicolumn{9}{l}{References: (c). \cite{ara13}; (d). \cite{mil20b}; (e). \cite{xu20c}; (f). \cite{mil20c}; (g). \cite{xu20a}; (h). \cite{mil20a};}\\
		\multicolumn{9}{l}{(i). \cite{bor12,bor13}, \cite{xu19}; (j). \cite{cha15b}, \cite{xu18}; (k). \cite{xu18}; (l). \cite{mil18}}\\
		\multicolumn{9}{l}{(m). Lower limits in columns 5-9 with associated 1-$\sigma$ uncertainty incorporated within the limit.}\\
		\multicolumn{9}{l}{(n). Lower limits in columns 5-9 with associated 1-$\sigma$ uncertainties.}\\
		\multicolumn{9}{l}{(o). Upper limits in columns 5-9 with associated 1-$\sigma$ uncertainties.}\\
	\end{tabular}
\end{table*}




\bsp	
\label{lastpage}
\end{document}